\documentclass{ieeeaccess}
\usepackage{cite}
\usepackage{amsmath,amssymb,amsfonts}
\usepackage{algorithmic}
\usepackage{graphicx}
\usepackage{color,soul}
\usepackage{textcomp}
\usepackage{subfig}
\def\BibTeX{{\rm B\kern-.05em{\sc i\kern-.025em b}\kern-.08em
    T\kern-.1667em\lower.7ex\hbox{E}\kern-.125emX}}
\begin{document}
\history{Date of publication xxxx 00, 0000, date of current version xxxx 00, 0000.}
\doi{10.1109/ACCESS.2017.DOI}

\title{A Graphene Field-Effect Transistor Based Analogue Phase Shifter for High-Frequency Applications}
\author{\uppercase{A. Medina-Rull\authorrefmark{1},}
\uppercase{F. Pasadas\authorrefmark{2},}
\uppercase{E. G. Marin\authorrefmark{1},}
\uppercase{A. Toral-Lopez\authorrefmark{1},}
\uppercase{J. Cuesta\authorrefmark{1},}
\uppercase{A. Godoy\authorrefmark{1},}
\uppercase{D. Jim\'enez\authorrefmark{2} and F. G. Ruiz\authorrefmark{1}.}}
\address[1]{PEARL Laboratory, Department of Electronics, Faculty of Science, University of Granada. Av. Fuentenueva S/N, 18071 Granada, Spain}
\address[2]{Departament d'Enginyeria Electr\`onica, Escola d'Enginyeria, Universitat Aut\`onoma de Barcelona, Campus UAB, C/ de les Sitges s/n, 08193 Cerdanyola, Spain}


\markboth
{Author \headeretal: Preparation of Papers for IEEE TRANSACTIONS and JOURNALS}
{Author \headeretal: Preparation of Papers for IEEE TRANSACTIONS and JOURNALS}

\corresp{Corresponding author: A. Medina-Rull (e-mail: amrull@ugr.es).}

\begin{abstract}
We present a graphene-based phase shifter for radio-frequency (RF) phase-array applications. The core of the designed phase-shifting system consists of a graphene field-effect transistor (GFET) used in a common source amplifier configuration. The phase of the RF signal is controlled by exploiting the quantum capacitance of graphene and its dependence on the terminal transistor biases. In particular, by independently tuning the applied gate-to-source and drain-to-source biases, we observe that the phase of the signal, in the super-high frequency band, can be varied nearly $200^\circ$ with a constant gain of $2.5\,$dB. Additionally, if only the gate bias is used as control signal, and the drain is biased linearly dependent on the former (i.e., in a completely analogue operation), a phase shift of $~85^\circ$ can be achieved making use of just one transistor and keeping a gain of $0\,$dB with a maximum variation of $1.3\,$dB. The latter design can be improved by applying a balanced branch-line configuration showing to be competitive against other state-of-the-art phase shifters. This work paves the way towards the exploitation of graphene technology to become the core of active analogue phase shifters for high-frequency operation.
\end{abstract}


\titlepgskip=-15pt

\maketitle

\section{Introduction}
\label{sec:introduction}
\PARstart{I}{n} the last few years the number of applications where graphene is involved has increased drastically in almost every field of electronics. Its use in flexible electronics \cite{li2017self, petrone2014graphene} along with its intrinsic material properties \cite{neto2009electronic, lee2008measurement, wang2011compact, meric2013graphene, wu2013graphene, elias2011dirac, pop2013thermal, balandin2011thermal}, have postulated it as one of the main candidates to play a leading role in the future of the industry. However, due to the absence of band gap in graphene and its consequent inability (at the device level) to be effectively turned off, this progress has been especially notorious in the field of RF electronics. Some examples of the main advancements can be found among radio-frequency (RF) power detection applications \cite{shayganHighPerformanceMetal2017}, high-frequency (HF) transmission lines \cite{yangImprovingRadioFrequency2019}, RF low power applications \cite{meleGrapheneFETsBased2018}, fifth-generation (5G) antenna arrays \cite{sadonGrapheneArrayAntenna2017}, or printed sensing applications for the Internet of Things (IoT) \cite{tiwariHANDBOOKGRAPHENEMATERIALS2018}. However, in the RF field, there are still some electronics components that indeed play an essential role in multiple communication systems embedded in radars or satellites \cite{kim4BitPassivePhase2009, shinohara2003solar, campbellCompact5bitPhaseshifter2000}, that remain unexplored. One notorious case corresponds to phase shifters, elements of paramount importance in order to control and direct the main radiation lobe of antenna arrays. In particular, this ability of the arrays allows for a reduction in power consumption and an overall improvement of the signal-to-noise ratio (SNR) of the antenna \cite{balanis2016antenna}, both crucial features in communication systems.

Although phase shifters are well-known RF systems, the design of purely analogue architectures, advantageous due to their higher precision and reduced complexity, has been technologically limited to two main approaches. First, the use of varactors as periodical loads of transmission lines, so to modify the equivalent circuit capacitance with a control voltage \cite{nagraDistributedAnalogPhase1999, nagraMonolithicGaAsPhase1999, xuCoupledLineBasedCoupler2019, ellingerVaractorloadedTransmissionlinePhase2003}. However, due to its passive nature, these topologies will always present some insertion losses (IL). Second, the implementation of transistor-based architectures either in all-pass filter configuration with a flat amplitude passing band and a voltage-dependent phase \cite{ulusoyTunableDifferentialAllPass2011}, or emulating the varactor structure by employing high electron mobility transistors \cite{viveirosTunableAllpassMMIC2002}.

An alternative strategy is getting the phase variation through a quadrature amplitude modulation (QAM) technique, where the so-called in-phase/quadrature (I/Q) signals (with a 90$^\circ$ phase difference between each other) feed two voltage gain amplifiers (VGAs), and are later added up. The resulting phase shift is controlled by changing the relative amplitudes of the I/O signals \cite{chuaGHzProgrammableAnalog1998, koh13muCMOSPhase2007, yu60GHzPhase2010, yongFullyIntegratedSBand2009, kimImprovedWidebandAllPass2012, pepeTwoMmWaveVector2017}. This strategy results in quite complex circuits (including the I/Q generator, VGAs, and signal adders) as well as a digital control system.


However, in spite of the increasing number of effective prototypes of graphene RF devices such as graphene field-effect transistors (GFETs) \cite{pandey2018all}, as well as one-dimensional (1D) flexible RF diodes \cite{wang2019flexible},
none of them have already been explored for the aforementioned purpose.
In this context, we propose a bias-controlled analogue phase shifter based on a GFET by taking advantage of the possibility of tuning the graphene quantum capacitance with the FET terminal biases thanks to its low density of states around the Dirac point \cite{xiaMeasurementQuantumCapacitance2009}. Not only is the use of graphene-based technology for this application novel and relevant, but also the fact that the phase shift can be controlled solely by an analogue signal without impacting on its gain. In this regard, the proposed phase shifter architecture consists of only one device and the role of the control signal is played by the gate bias with the drain bias linearly depending on the former. The proposed design achieves a $85^\circ$ phase shift, keeping a gain of $0\,$dB with a maximum variation of $1.3\,$dB. In order to reduce the source mismatch, the design is improved by applying a balanced branch-line amplifier configuration providing return losses higher than $30\,$dB. The performance and main figures of merit (FoMs) of the proposed graphene-based phase shifters are compared against the state-of-the-art with promising results.

\section{Graphene FET as phase shifter}
The design and analysis of an RF phase shifter founded in graphene require a physics-based description of the electrical behavior of a GFET at a compact and analytical level suitable for standard circuit simulators. To this purpose, we employ the large-signal model implemented in Verilog-A by some of the authors \cite{pasadasLargeSignalModelGraphene2016b}, embedding it into Keysight$^\copyright$ Advanced Design System.
This GFET compact model  has been thoroughly validated in \cite{pasadasLargeSignalModelGraphene2016} by the assessment of the DC characteristics, transient dynamics, and frequency response of a variety of graphene-based circuits such as a HF voltage amplifier \cite{hanHighFrequencyGrapheneVoltage2011}, a high-performance frequency doubler \cite{wangHighperformanceTopgateGraphene2010}, a subharmonic mixer \cite{habibpourSubharmonicGrapheneFET2012}, and a multiplier phase detector \cite{yangGrapheneAmbipolarMultiplier2011} showing a very good agreement between measurements and simulations.

In order to proceed with the device-level analysis, it is important to first introduce the graphene technology parameters considered within the design. Table \ref{tab: GFET_params} summarizes them, where $T$ is the temperature; $\mu$ represents the effective carrier mobility; $V_{\rm G0}$ is the gate offset voltage; $\Delta$ is the inhomogeneity of the electrostatic potential due to electron-hole puddles; $W$ and $L$ are the channel width and length, respectively; and $C_{\rm ox}$ is the oxide capacitance per unit area.

\begin{table}[!b]
\renewcommand{\arraystretch}{1.3}
\caption{{\small Graphene technology parameters used in the design of the analogue phase shifter.}}
\label{tab: GFET_params}
\centering
\begin{tabular}{c|c|c|c} \hline \hline 
Variable & Value & Variable & Value \\ \hline
 $T$ & 300 K & $L$ & 1 $\mu$m\\
 $\mu$ & 2000 cm$^2$/Vs & $W$ & 1 $\mu$m\\
 $V_{\rm G0}$ & 0 V & $C_{\rm ox}$ & 0.11 pF/$\mu$m$^2$\\
 $\Delta$ & 0.1 eV & &\\ \hline \hline
\end{tabular} 
\end{table}

\begin{figure}
     	 \centering
		 \subfloat{\label{fig:capacitance_vgs}\includegraphics[width=8.8cm]{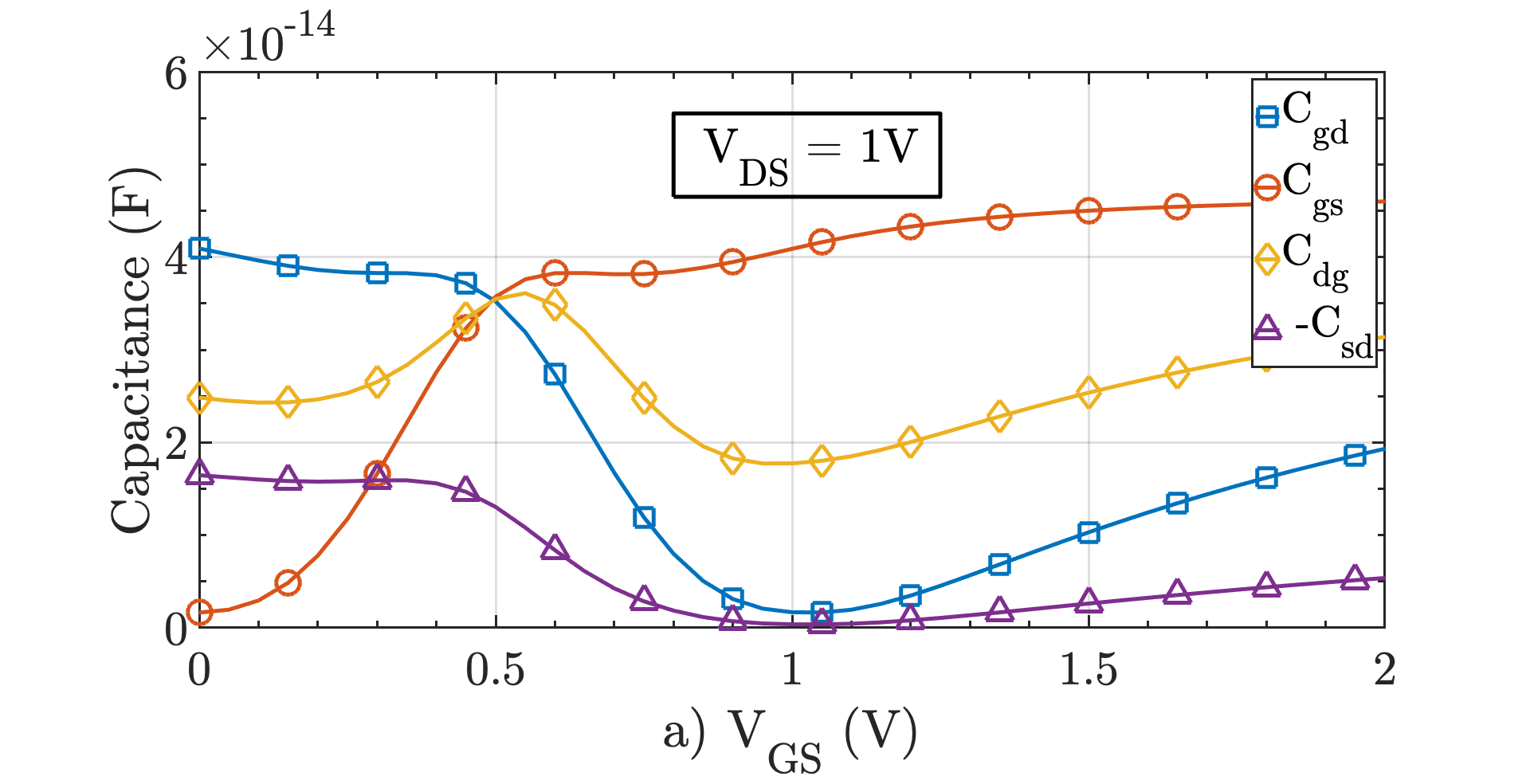}}         
         
         \subfloat{\label{fig:capacitance_vds}\includegraphics[width=8.8cm]{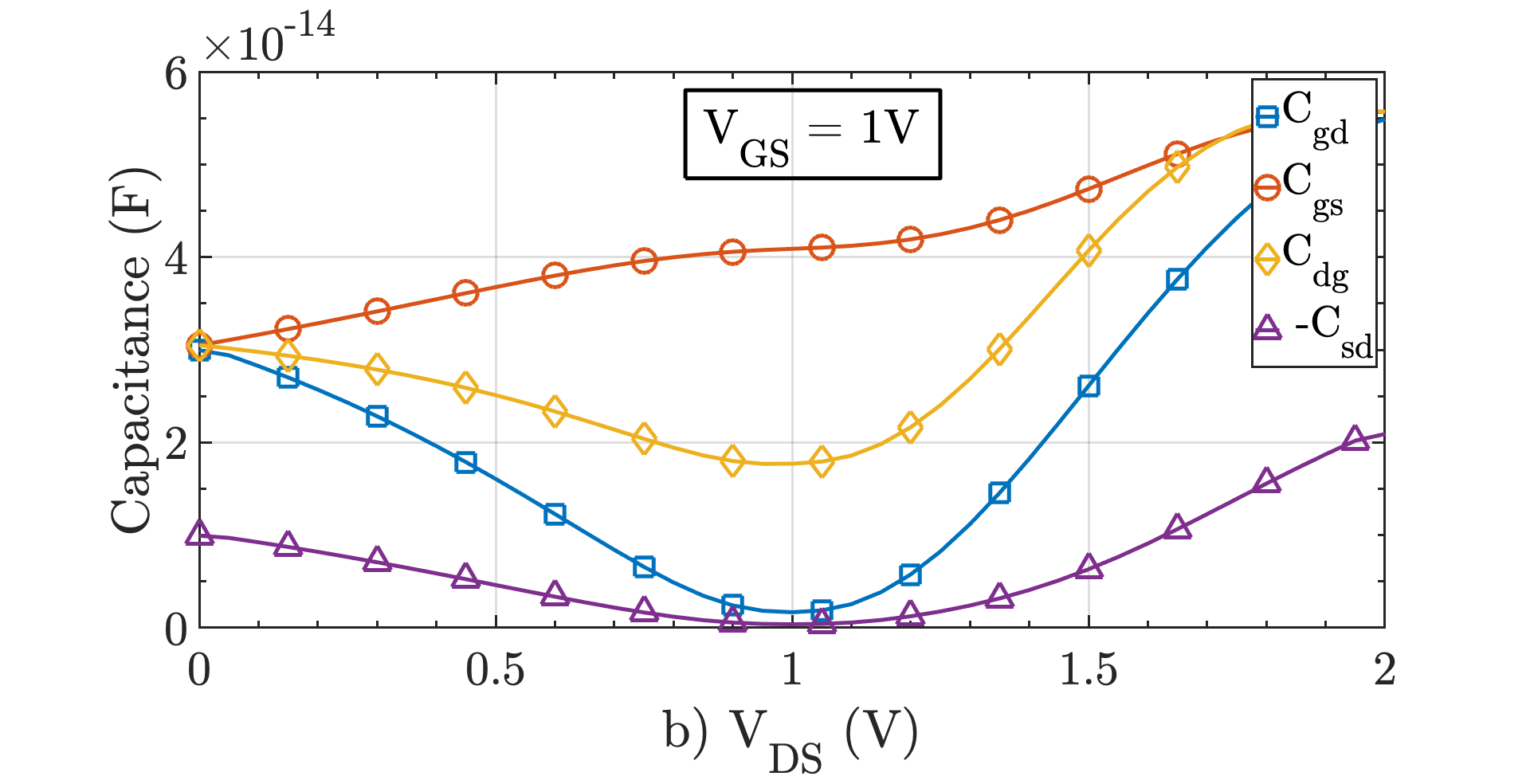}}
         
     \caption{{\small Intrinsic capacitances $C_{\rm gs}$ (red circles), $C_{\rm gd}$ (blue squares), $C_{\rm sd}$ (yellow diamonds) and $C_{\rm dg}$ (purple triangles) of the GFET employing the technology summarized in Table \ref{tab: GFET_params} versus (a) gate bias and (b) drain bias.}}
     \label{fig:capacitance}
\end{figure}

The inspirational property of a GFET that postulates it as a candidate to be the core of an active analogue phase shifter is the bias-tunable quantum capacitance originated by the reduced density of states of graphene around the Dirac point \cite{xiaMeasurementQuantumCapacitance2009}. To take advantage of this inherent property, the graphene quantum capacitance, $C_{\rm q}$, has to be dominant over the gate geometrical oxide capacitance, $C_{\rm {ox}} = \epsilon_0\epsilon_{\rm ox}/t_{\rm ox}$. In a metal-insulator-graphene structure $C_{\rm ox}$ and $C_{\rm q}$ are working in series \cite{moldovanGrapheneQuantumCapacitors2016}, and therefore, achieving a design with $C_{\rm ox} \gg C_{\rm q}$ allows to leverage the $C_{\rm q}$ tunability. This effect can be observed by analyzing the intrinsic device capacitances ($C_{ij}$) of the GFET which relate the incremental charge ($\Delta Q_i$) at a terminal $i$ with a varying voltage ($\Delta V_j$) applied to a terminal $j$ assuming that the voltage at all the other terminals remains constant \cite{pasadasLargeSignalModelGraphene2016b}, 

\begin{equation}
C_{ij} = \begin{cases} -\frac{\partial Q_i}{\partial V_j}, & i \neq j \\ \frac{\partial Q_i}{\partial V_j}, & i = j  \end{cases}
\label{eq:c_ij}
\end{equation}

\noindent where $i$ and $j$ stand for $g$ (gate), $d$ (drain), and $s$ (source) respectively. The dynamic regime of a three-terminal GFET can be described by just four out of nine intrinsic capacitances \cite{toral2019device, pasadas2019large}, $C_{ij}$ in Eq. (\ref{eq:c_ij}). In order to illustrate the device capacitive tunability with terminal biases, Fig. \ref{fig:capacitance} shows the gate ($V_\text{GS}$) and drain ($V_\text{DS}$) bias dependences of the selected set of capacitances, namely $C_{\rm gs}$, $C_{\rm gd}$, $C_{\rm sd}$ and $C_{\rm dg}$, considering the device technology described in Table \ref{tab: GFET_params}. As can be observed, all intrinsic capacitances show large variations in the selected range of bias proving that, due to the $C_{\rm q}$ tunability, using a control signal based on $V_\text{GS}$ and/or $V_\text{DS}$ can be eventually exploited for phase shifting operation through the variation of the capacitive response of the device.

To the purpose of selecting the RF band of operation, a bare and quick estimation of the RF performance limits of the GFET technology can be achieved by calculating the cut-off frequency, $f_T$, and maximum oscillation frequency, $f_{\rm max}$ \cite{pasadasSmallSignalModel2DMaterial2017, toral2019gfet}. In particular, the expected RF FoMs for the technology described in Table \ref{tab: GFET_params} are  $f_T = 18.9\,$GHz and $f_{\rm max} = 25.1\,$GHz at $V_\text{GS} = V_\text{DS} = 1\,$V. These values have been estimated considering that the source and drain metal-graphene contact resistances are limited to $R_s\cdot W = R_d\cdot W = 100\,\Omega\mu$m \cite{giubileoRoleContactResistance2017, moonUltralowResistanceOhmic2012, anziUltralowContactResistance2018}, respectively; and the gate resistance is $R_g\cdot L = 5\,\Omega\mu$m.
As a rule of thumb, the operating frequency should be lower than 20\% of $f_{\rm max}$ so as to guarantee sufficient power gain. In order to fulfill this requirement, we have opted for a design within the S-band of the spectrum. Specifically, the chosen operating frequency of the phase shifter is $3\,$GHz. Nevertheless, future improvements of the GFET technology would allow our design procedure to be extended beyond this frequency range.

\section{Phase shifter graphene circuit}

\begin{figure} [b]
 	\centering
 	\includegraphics[width=8.8cm]{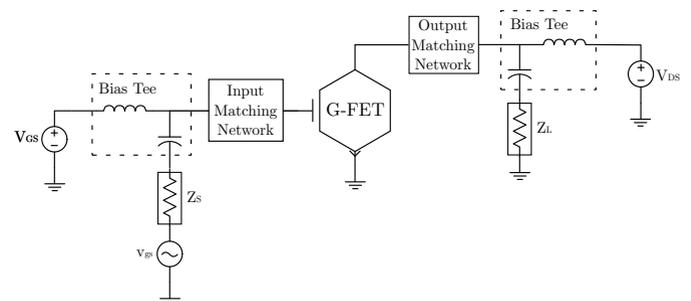}
 	\caption{{\small {Schematic of the phase shifter.} The GFET is used as the active element. IMN and OMN allow to maximize the power transfer from the source to the load and, at the same time, minimize signal reflection from the load. Bias tees at both input and output ports are considered, each one consisting of an ideal capacitor to allow the AC through but uncoupling the DC, and an ideal inductor to allow the DC through but uncoupling the AC signal.}}
 	\label{fig:Fig1}
 \end{figure}

As it is known, an analogue phase shifter is expected to produce a phase shift in the output with respect to the input, as dictated by a control signal, while the amplitude of the output  is minimally attenuated by a constant factor. In order to implement this concept using graphene, we propose a GFET operating in common-source (CS) configuration, thus forming a two-port network. As shown in Fig. \ref{fig:Fig1}, the RF signal and DC biases are combined by using bias tees consisting of L/C networks which properly block the AC/DC components, respectively. Source and load impedances, $Z_S$ and $Z_L$ respectively, are assumed equal to the characteristic impedance, set to $Z_0$ = $50\,\Omega$. In order to achieve a good power transfer, two matching networks are employed; while to the purpose of gaining stability, a shunt resistor of $1.65\,$k$\Omega$ is added to the gate of the GFET even though this will entail some gain losses. Unconditional stability is achieved for $V_\text{GS}$ and $V_\text{DS} = 1\,$V, which allows us to calculate the reflection coefficients $\Gamma_S$ and $\Gamma_L$ for the maximum available gain (MAG). Input and output matching networks (IMN and OMN, respectively) are designed to simultaneously satisfy $\Gamma_{in} = \Gamma_S^*$ and $\Gamma_{out} = \Gamma_L^*$ so as to yield conjugate matching in both ports. Both matching networks are composed by a shunt capacitor ($C_{\rm IMN} = 465.26\,$fF, $C_{\rm OMN} = 55.13\,$fF) and a series inductor ($L_{\rm IMN} = 35.07\,$nH, $L_{\rm OMN} = 37.32\,$nH). The IMN is configured in a C-L topology while the OMN is configured in a L-C topology. It is important to note that the lumped components here used are assumed to be ideal. Tolerance and quality factor (Q) issues associated with them could affect the design and, therefore, they should be analyzed in detail for the integrated circuit technology employed in an eventual realization of the circuit.

In a phase controlled antenna array, the phase shifter feeds each element of the array in a way that the amplitude and phase difference of the input current at each element determine the shape and direction of the main lobe of radiation, respectively. Thus, for a proper array design, it is of utmost relevance to be able to select the direction of the main lobe (changing the relative phases between the input signals of the antennas), while keeping the shape of the radiation pattern unaltered (maintaining the signal amplitudes of all elements balanced). Therefore, in terms of the scattering ($S$) parameters, a two-port phase shifter feeding each antenna must be able to keep the magnitude of $S_{21}$ ($|S_{21}|$) constant while tuning its phase in a controlled way ($\phi_{21}$), where ports $1$ and $2$ of the system refer here to the gate-source and drain-source terminals, respectively. The rest of the $S$ parameters ($S_{11}$, $S_{12}$ and $S_{22}$) are also important to guarantee an acceptable power transfer from the input to the output and are addressed by the proper design of the IMN and OMN. 

In particular, IMN and OMN in Fig. \ref{fig:Fig1} are optimized in order to achieve a value of the matching coefficient (M) as high as possible. Generally, matching networks are designed for a single bias point, but in this case, both $V_\text{GS}$ and $V_\text{DS}$ of the GFET have to be changed so to enable the phase shifting while keeping a constant amplitude, so it is crucial to have a high M value for a large window of $V_\text{GS}$ or $V_\text{DS}$ combinations. In this regard, we have assessed that M is over 0.7 for the bias window under test.

\begin{figure}
         \centering
         \subfloat{\label{fig:absS21_isocurve}\includegraphics[width=8.8cm]{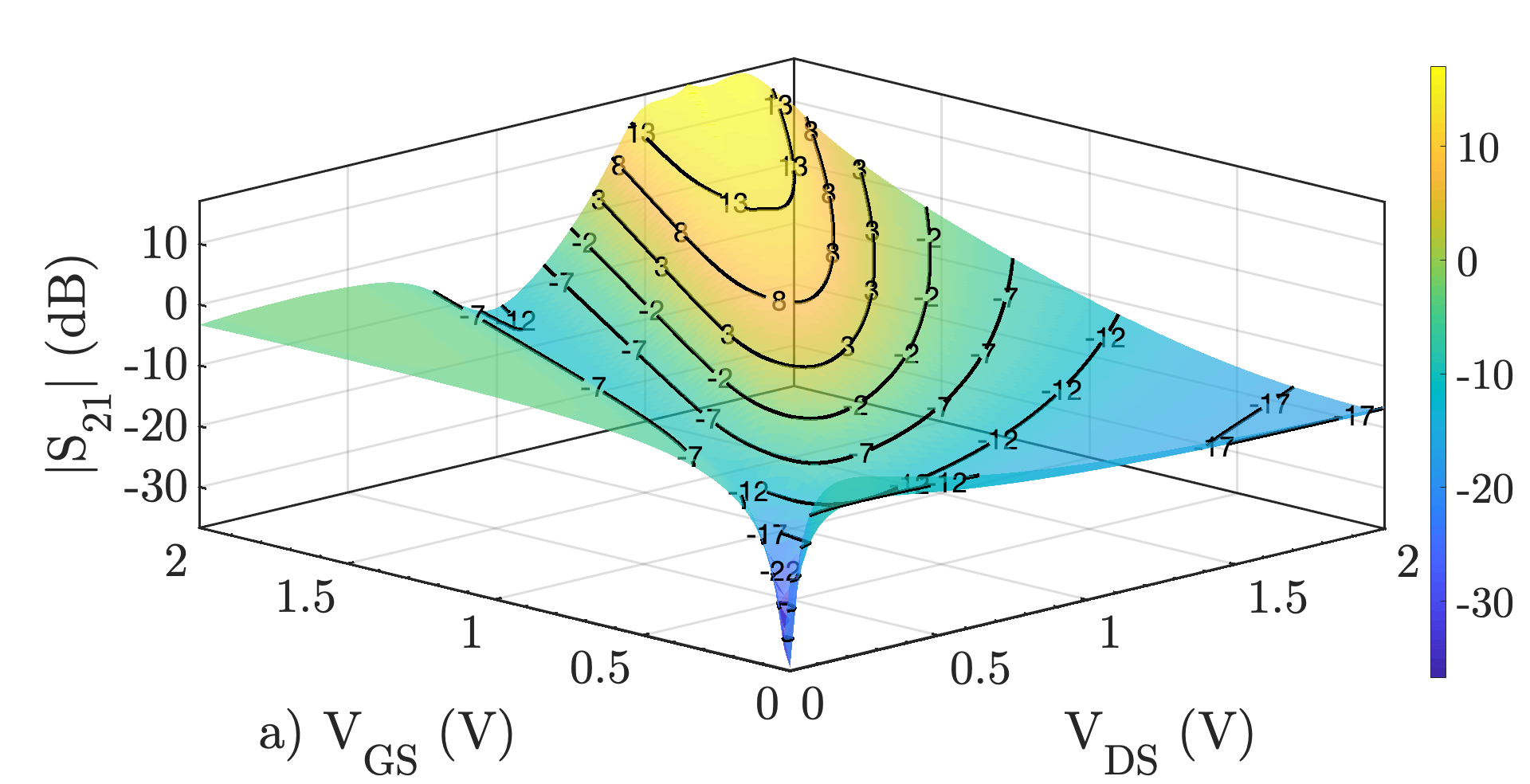}}
         
         \subfloat{\label{fig:phaseS21_isocurve}\includegraphics[width=8.8cm]{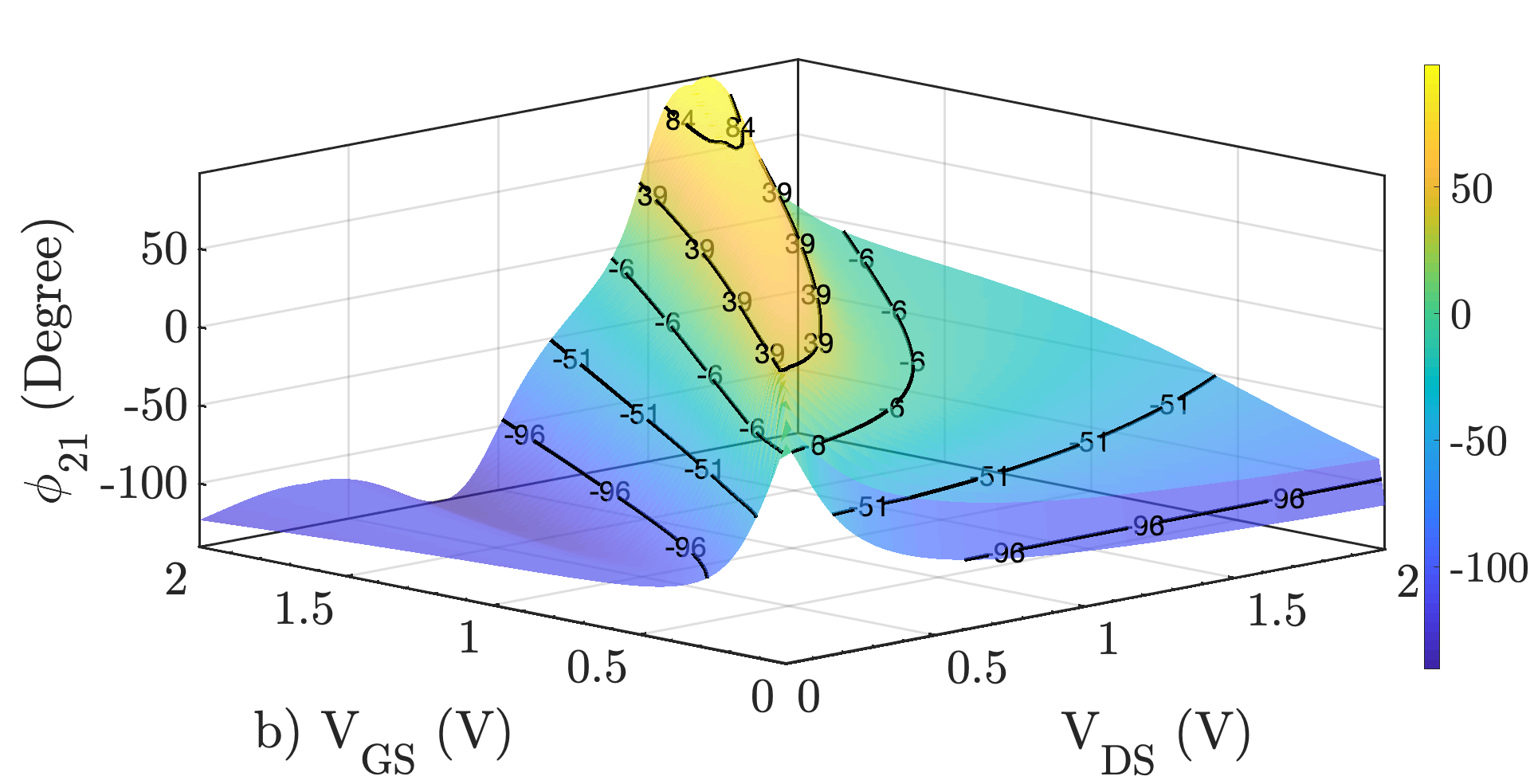}}
         
     \caption{{\small Isocurve plots of a) $|S_{21}|\,$(dB) and b) $\phi_{21}\,$(Degree) versus both $V_\text{GS}$ and $V_\text{DS}$.}}
     \label{fig:surface}
\end{figure}

In the design of Fig. \ref{fig:Fig1}, we expect that, by changing $V_\text{GS}$ or $V_\text{DS}$, the intrinsic capacitances of the GFET will vary, as shown in Fig. \ref{fig:capacitance_vgs}, and so will do $\phi_{21}$. The interest of the design is to keep at the same time a constant $|S_{21}|$. In order to better understand the $\phi_{21}$ and $|S_{21}|$ dependencies on the bias, Figs. \ref{fig:absS21_isocurve} and \ref{fig:phaseS21_isocurve} depict their corresponding isocurves as a function of $V_\text{DS}$ and $V_\text{GS}$.
As can be observed, both $|S_{21}|$ and $\phi_{21}$ show a strong dependence on $V_\text{DS}$ and $V_\text{GS}$ what can be exploited for the design of the phase shifter.
It should be highlighted that each isocurve of Figs. \ref{fig:absS21_isocurve} and \ref{fig:phaseS21_isocurve} provides a $V_\text{GS}-V_\text{DS}$ combination ensuring a constant $|S_{21}|$ and $\phi_{21}$. Moreover, the phase isocurves depict a different dependence on $V_\text{GS}-V_\text{DS}$ compared to amplitude isocurves, unveiling the possibility of applying a bias combination (i.e., a simultaneous variation of both ${V}_{\rm GS}$ and ${V}_{\rm DS}$) such that, harnessing the quantum capacitance tunability of graphene, would yield a constant amplitude while the phase is appropriately modified.

It is also interesting to note here that it would be possible to change the design technique and playing with the amplitude of the output signal ($|S_{21}|$) while maintaining a constant phase shift $\phi_{21}$, and this result would also be of notable interest as it would allow the radiation pattern to change while keeping the direction of the main lobe. 


 \begin{figure}
         \centering
         \subfloat{\label{fig:inset_stability}\includegraphics[width=8.8cm]{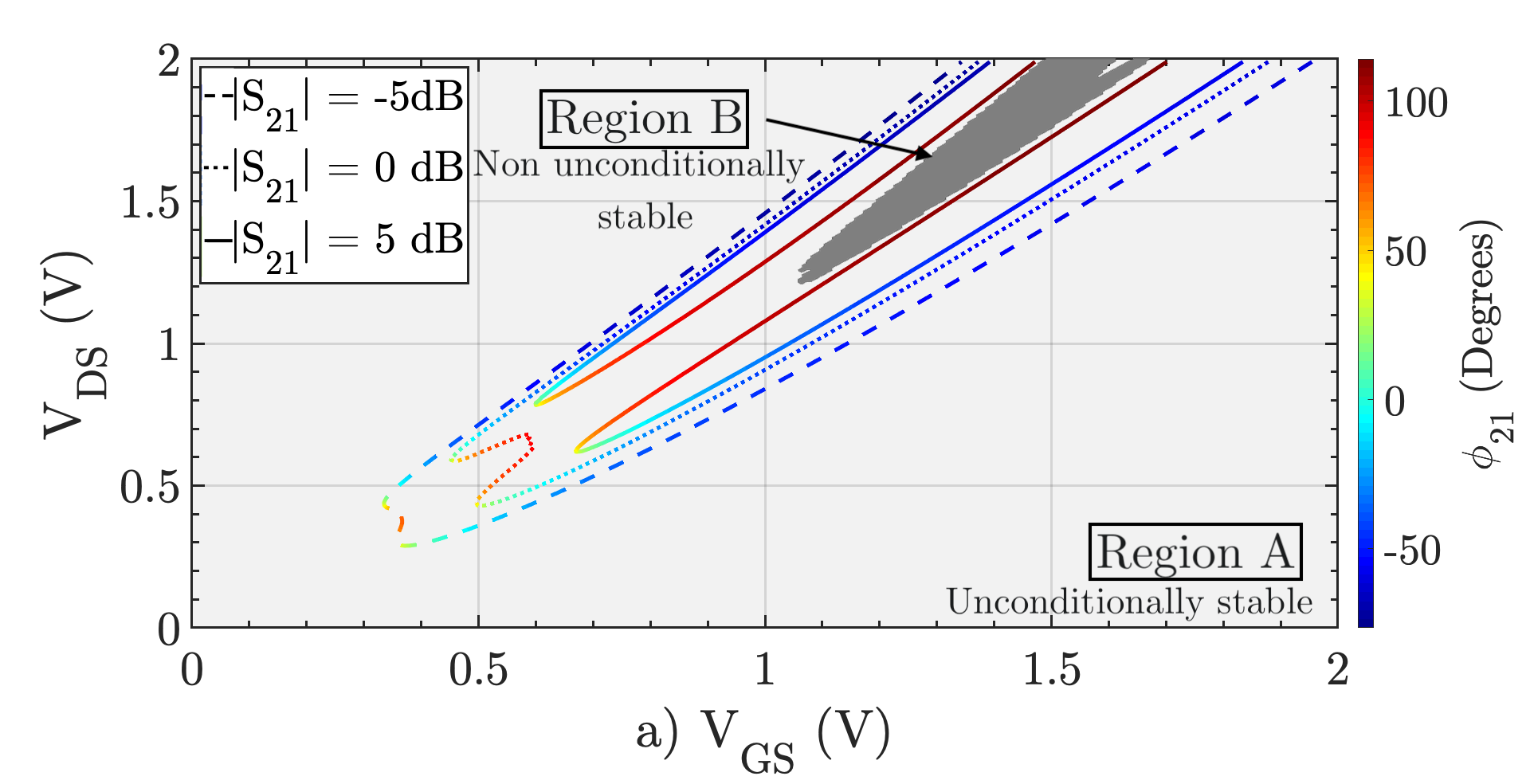}}
         
         \subfloat{\label{fig:phase_VGS}\includegraphics[width=8.8cm]{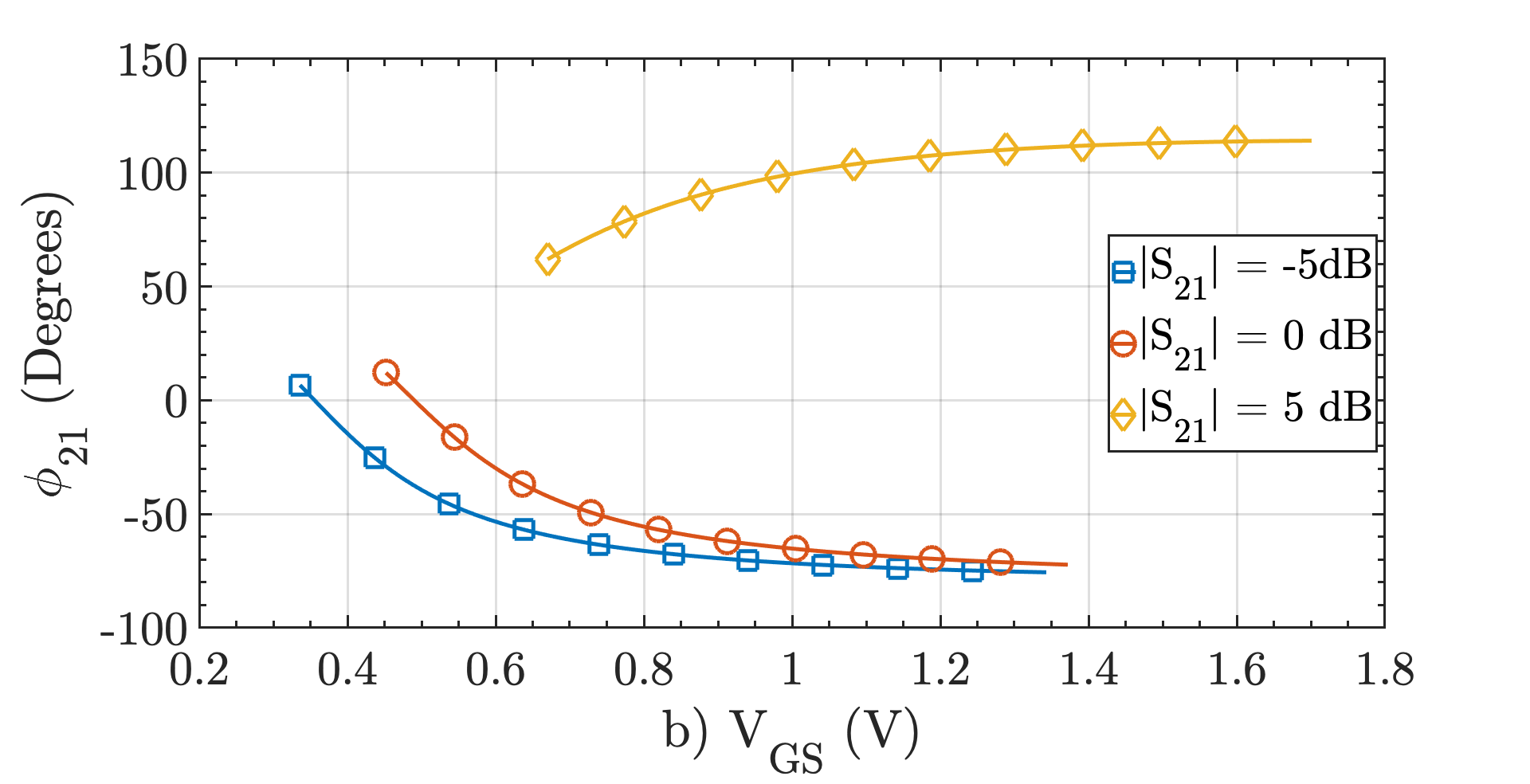}}
         
     \caption{{\small a) Bias dependence of the phase shift $\phi_{21}$ for three gain values $|S_{21}|$ = $-5\,$dB (dashed line), $0\,$dB (dotted line) and $5\,$dB (solid line). Bias combinations that do not guarantee unconditional stability for the device are coloured in dark grey, and are represented by region B. b) Gate bias dependence of the phase shift for the same three constant gain values by considering that the drain bias is simultaneously modified to mantain the selected $|S_{21}|$ value (analogue control).}}
     \label{fig:phase_vs_gain}
\end{figure}

Following on with the phase shifter design, Fig. \ref{fig:inset_stability} shows the $\phi_{21}$ variation (color scale) as a function of the bias combinations that keep a constant value of $|S_{21}|$: $-5\,$dB (dashed line), $0\,$dB (dotted line) and $5\,$dB (solid line). In order to ensure the unconditional stability of the circuit, the so-called K-$\Delta$ test \cite{pozarMicrowaveEngineering2012} is carried out. In that regard, Fig. \ref{fig:inset_stability} shows two different regions denoted as A (light grey) and B (dark grey). Region A represents the bias combinations where unconditional stability is achieved. As for Region B, it contains the bias points where the stability of the circuit cannot be assured, i.e., there are bias combinations inside Region B where either the stability is conditioned or the circuit is directly unstable. For this reason, we have chosen to restrict the design to Region A, ensuring that the circuit is unconditionally stable.

Using a purely digital control, i.e., allowing any possible $V_\text{GS}-V_\text{DS}$ combinations that provide a constant specific gain, would result in large phase shift ranges, e.g. $\Delta \phi_{21} \simeq 180^\circ$ keeping $|S_{21}| = 0\,$dB. 
If an analogue control is considered, i.e., a linear relation is forced between $V_\text{DS}$ and $V_\text{GS}$, the range of $\Delta \phi_{21}$ diminishes. Nevertheless, the analogue control would rely only on one signal, e.g., $V_\text{GS}$, and an extraordinary simple control circuit would be required (that may be implemented by a DC-DC converter, or in case efficiency is not a constraint, a simple voltage divider). This outstanding linear relation between both biases can be estimated for each $|S_{21}|$ isocurve by applying a standard linear regression and ensuring that the determination coefficient ($R^2$) is higher than $0.9999$. In this regard, Fig. \ref{fig:phase_VGS}) shows the performance of the analogue controlled phase shifter, demonstrating $\Delta \phi_{21}$ values higher than 50$^\circ$ for the three gain values considered and with a remarkable value of $\Delta \phi_{21}$ higher than $80^\circ$ for $|S_{21}| = 0\,$dB.

\begin{figure} [t]
   \centering
 	\includegraphics[width=8.8cm]{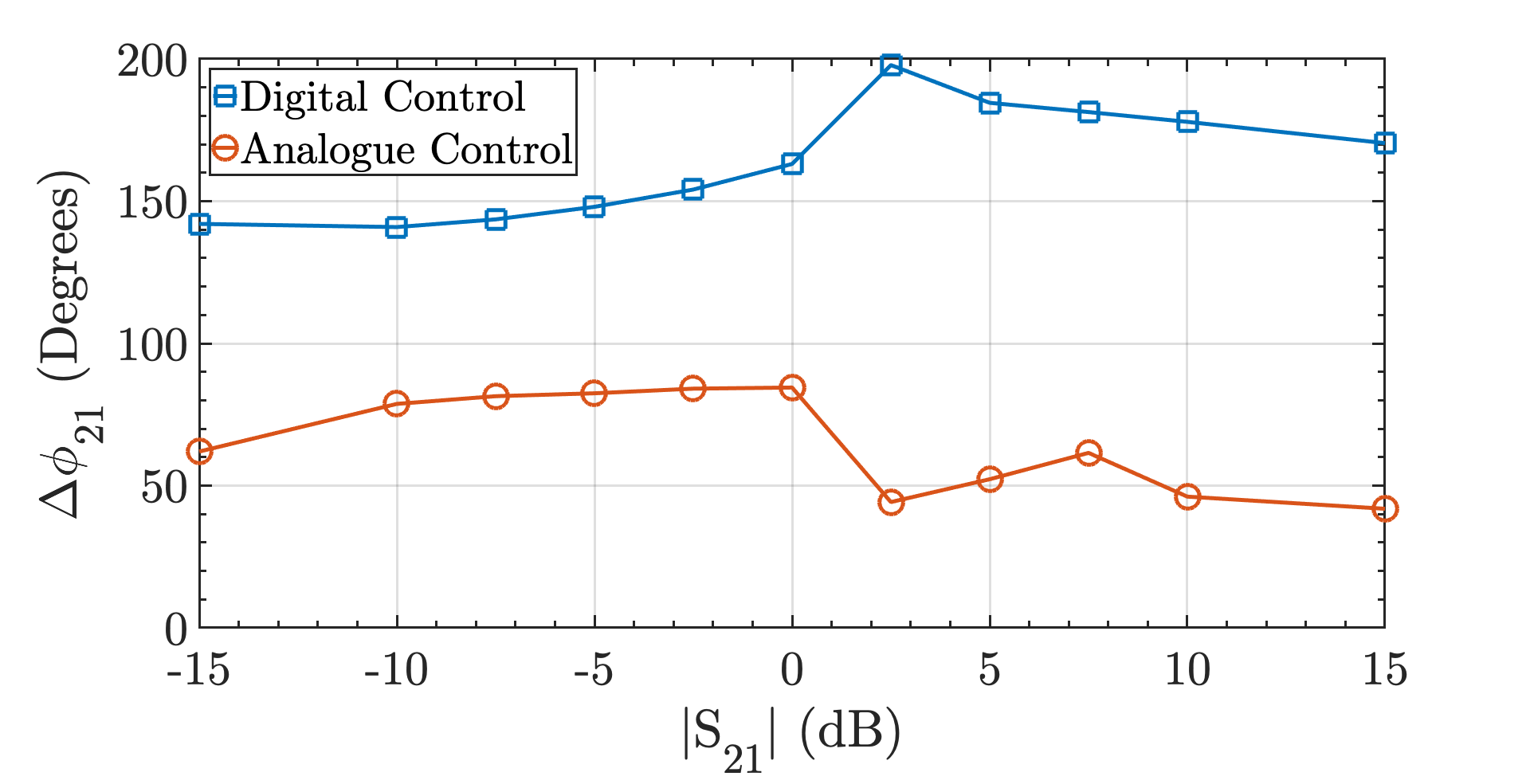}
 	\caption{{\small Maximum feasible phase shift range with digital (blue squares) and analogue (red circles) control versus $|S_{21}|$.}}
 	\label{fig:max_phaseshift}
 \end{figure}

As the maximum phase shift range $\Delta \phi_{21}$ depends on the gain value, we have evaluated it under two scenarios: (i) when the digital control is selected, and (ii) when the linear relation between $V_\text{GS}$ and $V_\text{DS}$ is assumed. The results are depicted in Fig. \ref{fig:max_phaseshift}, where $\Delta\phi_{21}$ is plotted for gains ranging from $-15\,$dB up to $15\,$dB, considering either a digital or an analogue control. According to Fig. \ref{fig:max_phaseshift}, $|S_{21}| = 0\,$dB happens to be in good trade-off between power gain and phase shift when designing an analogue phase shifter. It should be highlighted that it is possible to obtain a higher gain at the expense of losing some phase shift. This trade-off, however, can be balanced with the inclusion of an additional amplifier in the design of the phase shifter.


%

\begin{figure}
         \centering
         \subfloat{\label{fig:final_results}\includegraphics[width=8.8cm]{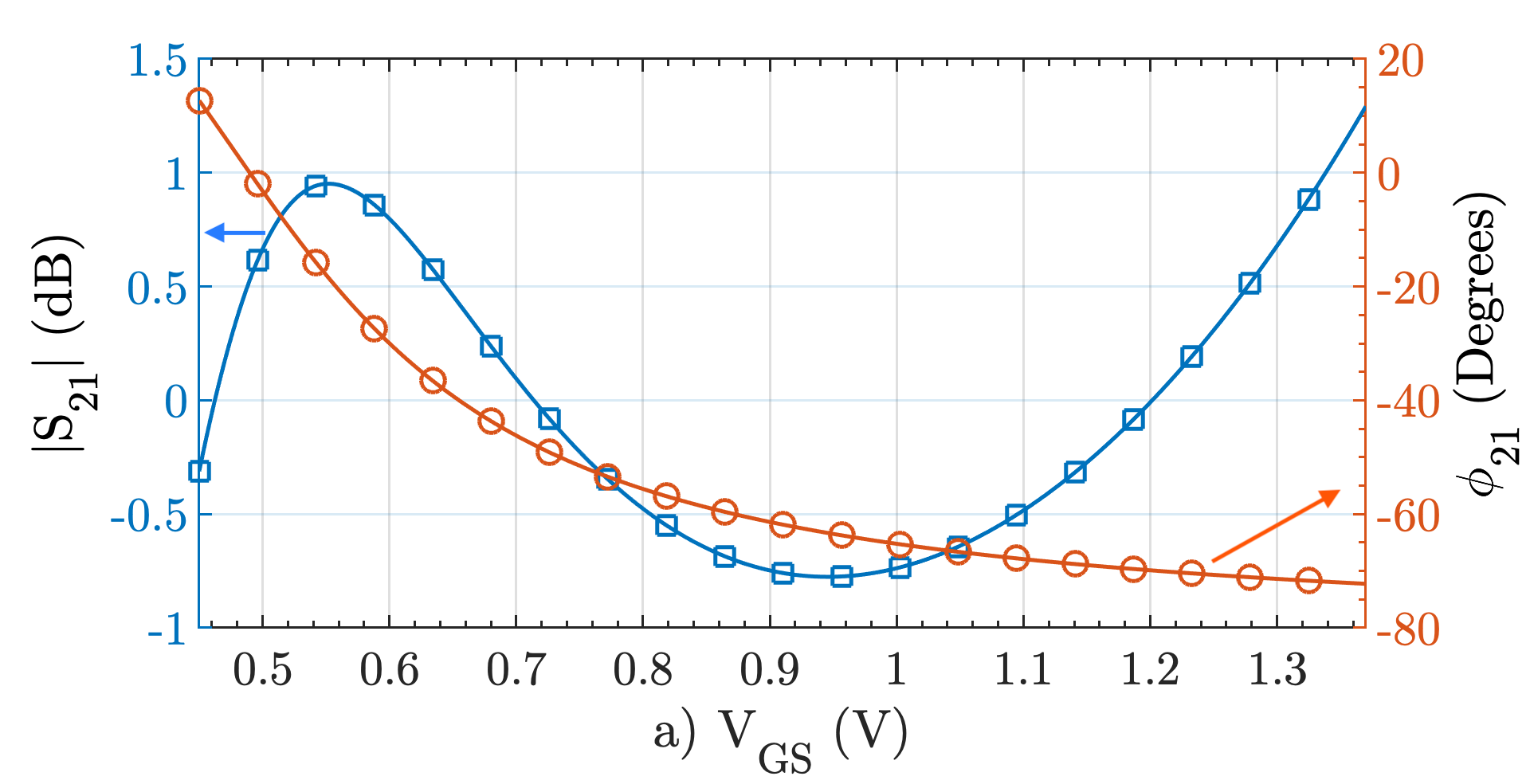}}
         
         \subfloat{\label{fig:linearity}\includegraphics[width=8.8cm]{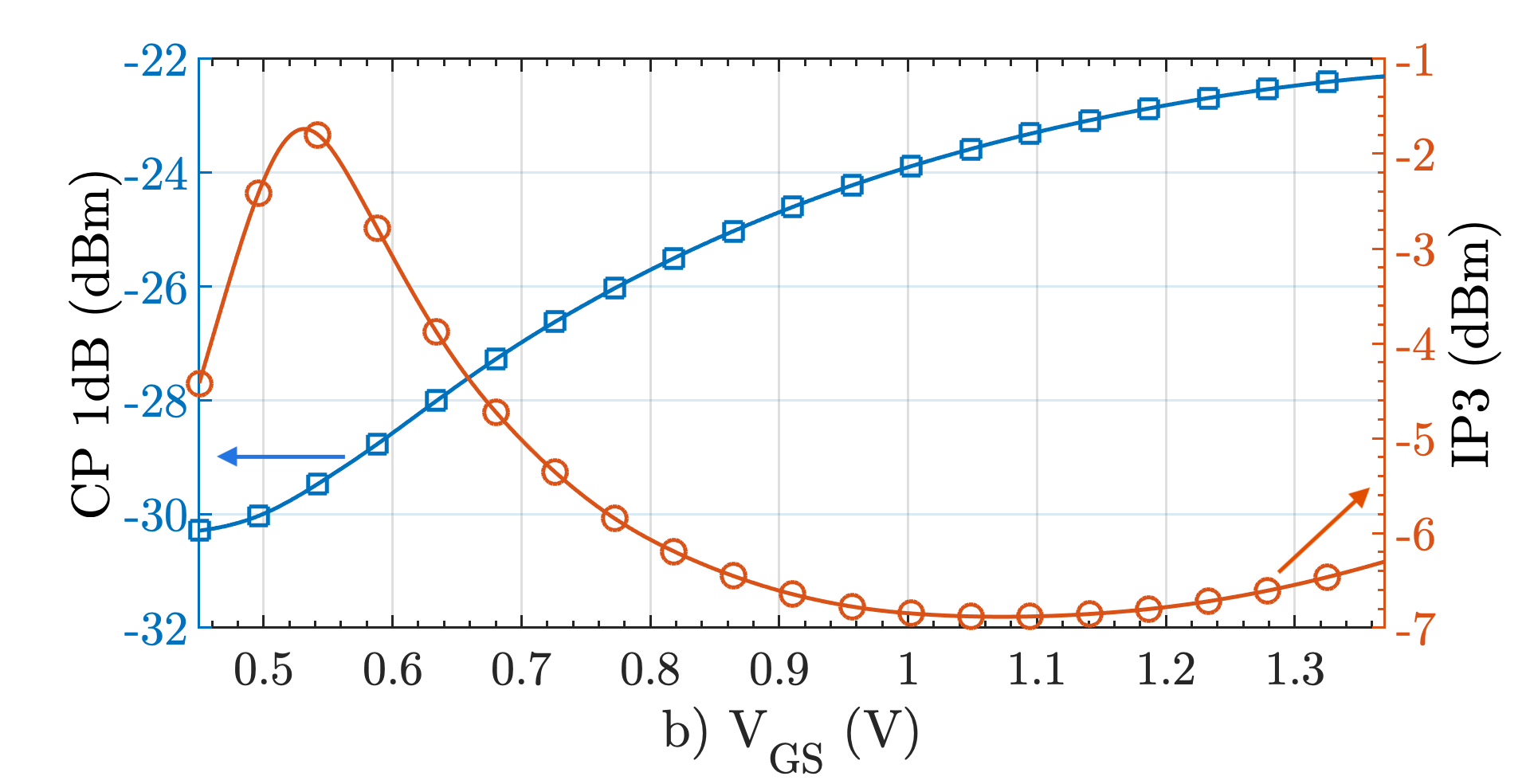}}
         
     \caption{{\small a) $|S_{21}|$ (blue squares) and $\phi_{21}$ (red circles) variation; and b) compression point at 1 dB (blue squares) and third order interception point (red circles) versus the analogue control provided by the gate bias.}}
     \label{fig:non_linearity}
\end{figure}

In order to evaluate the variations in $|S_{21}|$ that the purely analogue control of the phase shifter induces (due to the small depart from linearity of the actual $V_\text{GS}-V_\text{DS}$ combinations), Fig \ref{fig:final_results} depicts the outcome of the analogue controlled phase shifter for $|S_{21}| = 0\,$dB. As can be seen, when forcing the linear relation for $V_\text{GS}-V_\text{DS}$, still a $85^\circ$ phase shift range is achieved while satisfying a gain of 0 dB and a maximum variation of 1.3 dB of  $|S_{21}|$.

Fig. \ref{fig:linearity} completes the analysis of the analogue controlled phase shifter showing the compression point at $1\,$dB (CP $1\,$dB) and the third order interception point (IP3) as main FoMs to assess the linearity of the amplifier. The CP $1\,$dB is considerably low (lower than $-30$ dBm for the worst case), which limits the input power of the phase shifter to $-30\,$dBm. When using this device in reception applications, the signal may be free from any distortion as the power of the received signals in most of the wireless transmission protocols is typically lower than that value. However, the current design for the graphene phase shifter would be quite limited for transmission applications, and power stages should be added after it to provide enough power to the transmitted signal. In any case, due to the likely interest of using the proposed phase shifter as both, transmitter and receiver, a bidirectional configuration of the device is proposed in Appendix \ref{sec.bidirectional}. In spite of this, it is worth to mention that improvements in the performance of GFET technology are expected in the future \cite{bonmann2018graphene}, so that this non-linear behavior should be improved.

It is also important to mention that a frequency analysis of the proposed phase shifter has been carried out. The results show that the phase shifter in its current form is only suitable to work in narrowband applications. Further developments should be implemented in order to increase the bandwidth of the circuit, and in particular input and output wideband matching networks should be used.
 
We have carried out in Table 2 a comparison among different state-of-the-art topologies currently employed to implement phase shifters and the one proposed in this work. In terms of the IL, our proposal is the only one that is able to supply some gain to the signal, thanks to the use of an amplifier configuration based on the GFET. On the other hand, the main limitations of our design are: (i) return losses (RL) are still limited to poor values, and (ii) the range of phase shift is not the widest, although this is to a certain point balanced by the particular simplicity of the analogue control hardware.

\begin{table*}
\label{tab: comparison}
\caption{{\small Comparison among state-of-the-art phase shifter topologies.}}
\centering
\begin{tabular}{c|c|c|c|c|c|c} \hline \hline 
Shift Method & Control Type & Frequency (GHz) & IL (dB) & RL (dB) & $\Delta\phi$ (Degrees) & Reference\\ \hline
Switched Line & Digital & 13-18 & 2.7 & 22 & 349.3 & \cite{deyReliabilityAnalysisKuBand2015}\\
 Reflection type & Analogue & 2 & 1 & 13.4 & 385 & \cite{burdinDesignCompactReflectionType2015}\\
  Network type & Digital & 0.5-1 & 2.5 & 13 & 360 & \cite{tangDesignConsiderationsOctaveBand2010}\\
 Loaded Transmission Line & Analogue & 1 & 2 & 15 & 183 & \cite{linContinuouslyTunableTrueTimeDelay2019}\\ 
 GFET CS Amplifier & Digital & 3 & -2.5 & 0.9 & 197.9 & This work \\ 
 GFET CS Amplifier & Analogue & 3 & 0 & 0.4 & 84.5 & This work  \\ 
 Balanced Amplifier & Analogue & 3 & 0 & 30.4 & 85.5 & This work\\ \hline \hline
\end{tabular}
\end{table*}

\begin{figure} [h]
 \centering
 	\includegraphics[width=8.8cm]{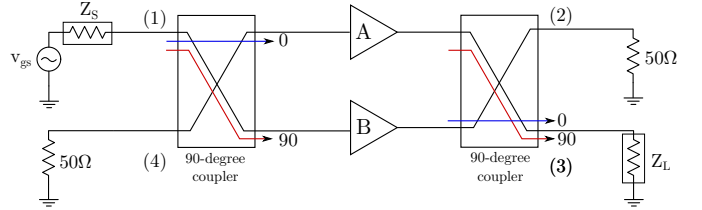}
 	\caption{{\small Schematic of the balanced amplifier. Two amplifiers, A and B, are used along with two $90^\circ$ hybrid couplers. The schematic of the amplifiers is shown in Fig. \ref{fig:Fig1}.}}
 	\label{fig:balanced_squematic}
 \end{figure}

With the aim of attaining a better performance in terms of return losses, we explore the feasibility of using a balanced configuration with a branch-line coupler. The attention paid to the achievement of flat insertion losses and a wide range of phase shift, may lead to a degradation of the standing-wave-ratio (SWR) at input and output ports, thus endangering the power generator as the reflected power could be too high. In our particular situation, we are continuously changing the bias of the device, and therefore operating the device in different bias points to those for which the matching networks were originally designed. This is the reason why the analysis of the return losses shown in Table 2 gives poor results in the worst case. 
The adoption of a balanced configuration solves this issue by using two hybrid couplers at input and output ports, along with two amplifiers \cite{gonzalezMicrowaveTransistorAmplifiers1997}. The schematic of the circuit designed to this purpose is depicted in Fig. \ref{fig:balanced_squematic}. The hybrid coupler is characterized by its scattering parameters, which can be ideally described as:

\begin{equation}
[S] = \frac{-1}{\sqrt{2}}
\begin{bmatrix}
0 & j & 1 & 0\\
j & 0 & 0 & 1\\
1 & 0 & 0 & j\\
0 & 1 & j & 0
\end{bmatrix}\label{eq.paramS_coupler}
\end{equation}

So that, if all ports of the coupler are matched, the power entering into port 1 is evenly divided between ports 2 and 3, with a 90$^\circ$ phase shift one with respect to the other. Port number 4 is isolated, so no power would be coupled to it \cite{pozarMicrowaveEngineering2012}.
With the proper analysis, the global parameters of the network $S_{11}$ and $S_{21}$ can be calculated by following Eqs. (\ref{eq.S11_balanced}) and (\ref{eq.S21_balanced}), respectively, where $a$ and $b$ subscripts denote the $S$-parameters of A and B amplifiers respectively. It is of special interest the case where both amplifiers are alike, as in this case $S_{11} = S_{22} = 0$ and $S_{21}$ is equal to the gain of one of the parallel branches of the coupler (with a phase shift of 90$^\circ$). This means that it is possible to get rid of any reflected power at the input port, with no gain losses, at the cost of a higher complexity of the circuit.

\begin{equation}
S_{11} = \frac{e^{-j\pi}}{2}(S_{11a} - S_{11b})\label{eq.S11_balanced}
\end{equation}
\begin{equation}
S_{21} = \frac{e^{-j\frac{\pi}{2}}}{2}(S_{21a} + S_{21b})\label{eq.S21_balanced}
\end{equation}

 \begin{figure} [h]
 \centering
 	\includegraphics[width=8.8cm]{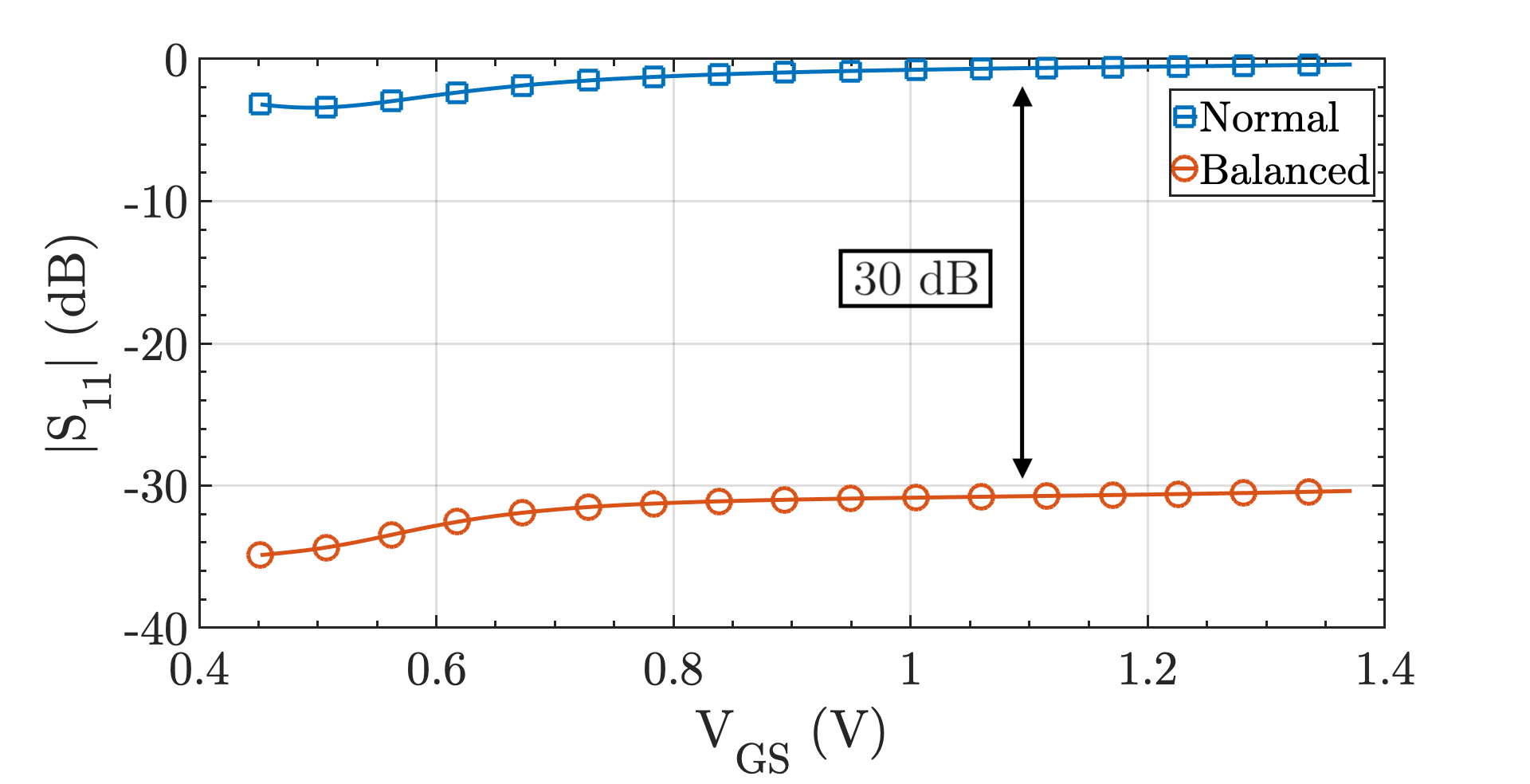}
 	\caption{{\small Comparison of the $|S_{11}|$ parameter (blue squares) of the former configuration presented in Fig. \ref{fig:Fig1} and of the balanced amplifier (red circles) presented in Fig. \ref{fig:balanced_squematic}.}}
 	\label{fig: S11_comparison}
 \end{figure}
 
 
Finally, Fig. \ref{fig: S11_comparison} compares $S_{11}$ as a function of the device bias for the single branch configuration (Fig. \ref{fig:Fig1}) and the alternative balanced configuration (Fig. \ref{fig:balanced_squematic}). As expected, the results show that as both branches of the design are identical, the reflection coefficient is strongly diminished, providing a reduction of more than $30\,$dB in $S_{11}$. Adopting this new configuration, the return losses are RL $= 30.4\,$dB for the worst case, which makes the balanced amplifier phase shifter topology comparable to the rest of the technologies considered in Table 2. Again, the use of this balanced configuration is possible thanks to the use of an amplifier configuration based on the GFET.

As for the phase shift range shown in Table 2, even though our solution yields a range which is one-fourth of other state-of-the-art devices, it does not preclude its use for numerous applications: many antenna arrays would need no more than 10$^\circ$ shift in their pointing, for which a controllable phase shift of 80$^\circ$ is enough.

A broader phase shift range would be easily achieved by cascading several balanced amplifiers. As the reflection parameter of the balanced amplifier configuration of the phase shifter is extremely low, the cascading would be successfully obtained and, therefore, a multi-stage phase shifter can be readily attained. Eventually, if we cascaded four of these balanced structures, a phase shift range of around 360$^\circ$ would be obtained, making our device fully competitive with other ones in terms of phase shift range.


\section{Conclusion}
This work presents a graphene-based phase shifter operating in the S-band, able to produce a phase shift on an input RF signal while maintaining a constant gain. Quantum capacitance tunability of graphene is leveraged in order to achieve this phase modulation, combined with an original design procedure. Phase shifts higher than $180^\circ$ are possible, as well as gains above 15$\,$dB, due to the amplifier configuration adopted with the GFET as the core element. Moreover, a completely analogue operation has been demonstrated achieving a phase shift of up to $84.5^\circ$ and keeping a maximum variation of $1.3\,$dB. The RL have been considerably increased by using a balanced configuration based on the amplifier topology employed. These results demonstrate the potential of graphene technology for the future development of improved high-frequency applications and in particular for analogue phase shifters.

\appendices

\section{Bidirectional Operation}\label{sec.bidirectional}
\begin{figure} [h]
 \centering
 	\includegraphics[width=8.8cm]{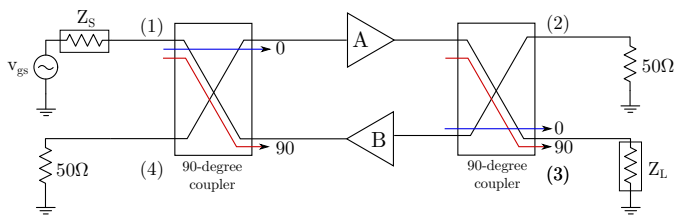}
 	\caption{{\small Schematic of the balanced amplifier in a bidirectional operation. In this case, amplifier B is mirrored vertically, so that its input is fed in port 3, and the output is extracted from port 1. The schematic of the amplifiers employed in this design is depicted in Fig. \ref{fig:Fig1}.}}
 	\label{fig:bidirectional_squematic}
 \end{figure}

This Appendix shows the results of employing the balanced amplifier phase shifter simultaneously as a transmitter and as a receiver. These calculations are of great importance, as they may be required for its application in phased-array techniques in reception and transmission, thus improving the global efficiency of the system not just in one direction but in both. This device is made by vertically mirroring amplifier B from the topology shown in Fig. \ref{fig:bidirectional_squematic} so that its input is fed in port 3 and the output is extracted from port 1. The final schematic is shown in Fig. \ref{fig:bidirectional_squematic}.

\begin{figure}[ht]
         \centering
         \subfloat{\label{fig:bidirectional_S12}\includegraphics[width=8.8cm]{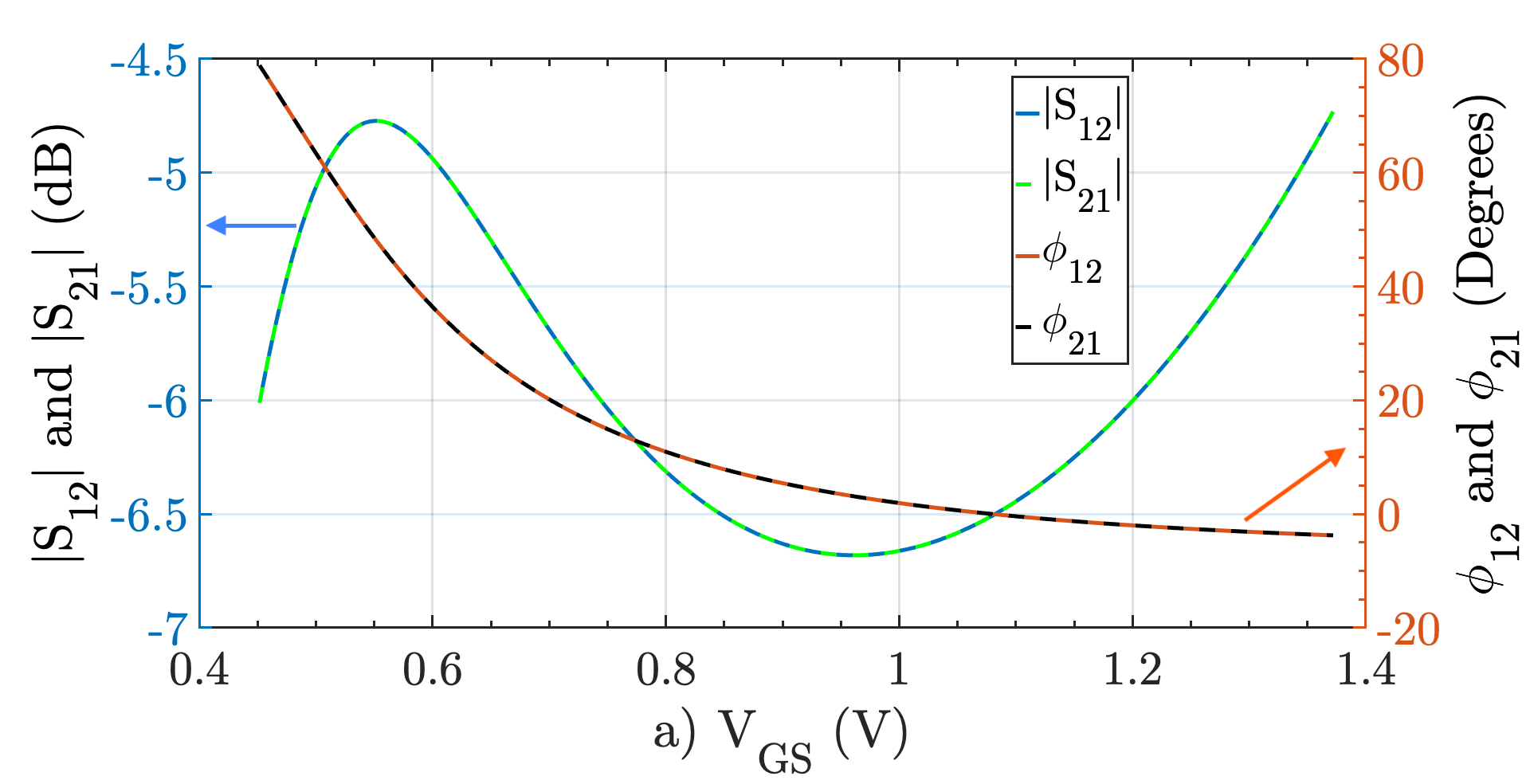}}
         %
     \hfill
         \subfloat{\label{fig:bidirectional_S22dB}\includegraphics[width=8.8cm]{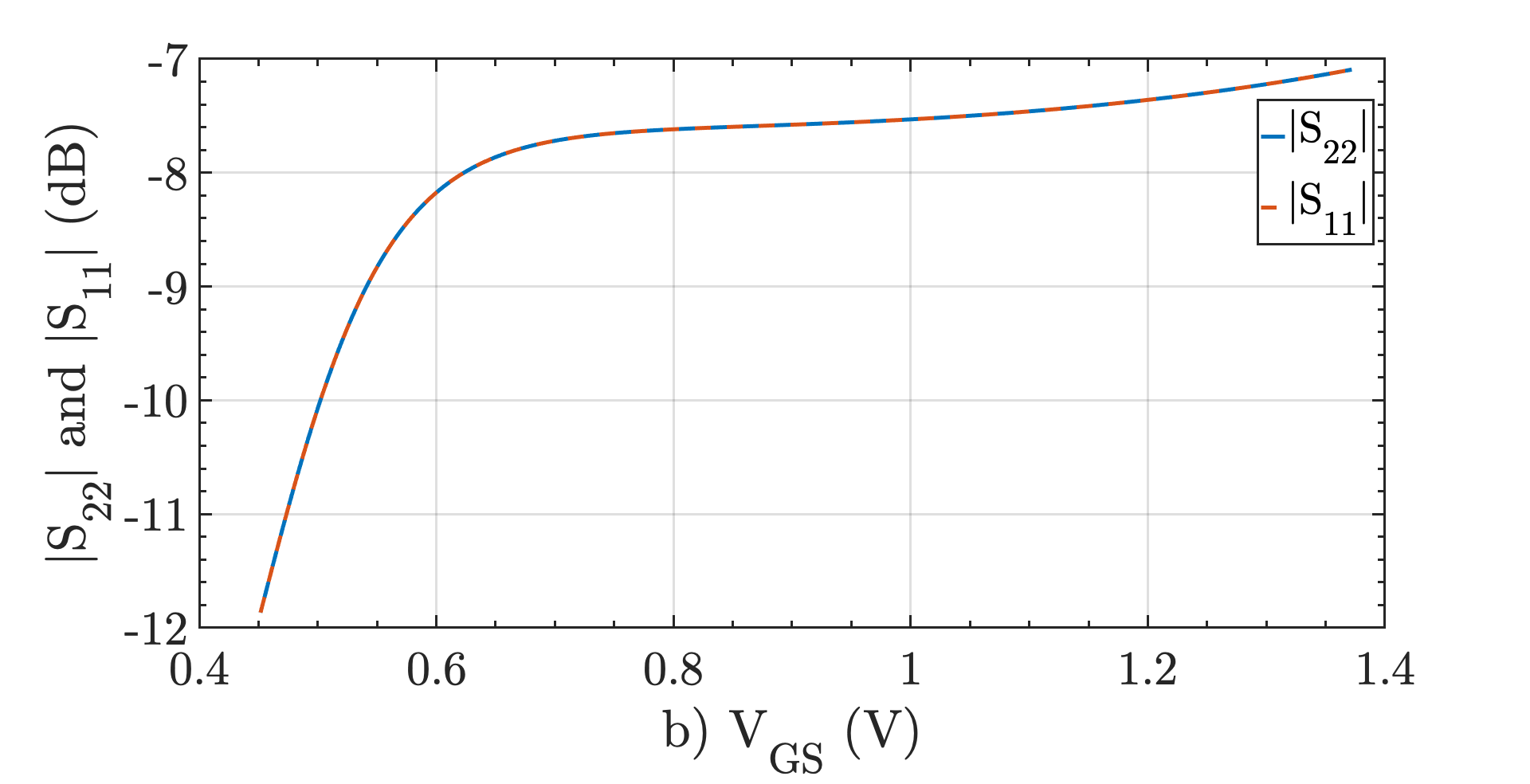}}
         %
     \caption{{\small a) $|S_{12}|$ and $\phi_{12}$ variation and b) $|S_{22}|$ and $|S_{11}|$ variation versus $V_\text{GS}$.}}
     \label{fig:bidirectional}
\end{figure}

In the bidirectional system, $S_{22} = S_{11}$ and $S_{12} = S_{21}$. As can be seen in Fig. \ref{fig:bidirectional_S12}, the shape of the amplitude and phase of the bidirectional amplifier is the same shown in Fig. \ref{fig:final_results}, but instead of $0\,$dB now we have a $-6\,$dB gain, which is still acceptable even though it means we are lossing some power. The phase is also shifted but the range of variation remains identical. The reflection parameter $|S_{22}| = |S_{11}|$ shows in Fig. \ref{fig:bidirectional_S22dB} values lower than $-7\,$dB for the worst case. Although this value is not optimal, it is still acceptable. In conclusion, we have demonstrated a system able to work as a phase shifter in both directions symetrically with a $-6\,$dB gain and RL = $7\,$dB for the worst case.

\section*{Acknowledgment}

This work was supported in part by the Ministerio de Ciencia, Innovaci\'on y Universidades, Agencia Estatal de Investigaci\'on (AEI), European Regional Developments Fund (ERDF/FEDER), under Project TEC2017-89955-P and Project EQC2018-004963-P (MINECO/AEI/FEDER), in part by the FEDER/Junta de Andaluc\'ia-Consejer\'ia de Econom\'ia y Conocimiento under Project B-RNM-375-UGR18; and in part by the European Commission through the Horizon 2020 Project Wearable Applications Enabled by Electronic Systems on Paper (WASP) under Contract 825213. The work of Enrique G. Marin was supported by the Juan de la Cierva Incorporaci\'on under Grant IJCI-2017-32297 (MINECO/AEI). This work was supported by the European Union\textquotesingle s Horizon 2020 Research and Innovation Programme under Grant GrapheneCore2 785219 and Grant GrapheneCore3 881603, in part by the Ministerio de Ciencia, Innovaci\'on y Universidades under Grant RTI2018-097876-B-C21(MCIU/AEI/FEDER, UE), in part by the European Regional Development Funds (ERDF) through the Programa Operatiu FEDER de Catalunya 2014-2020, with the support of the Secretaria d\textquotesingle Universitats i Recerca of the Departament d\textquotesingle Empresa i Coneixement of the Generalitat de Catalunya for emerging technology clusters to carry out valorization and transfer of research results, and in part by GraphCAT under Project 001-P-001702.

\bibliographystyle{IEEEtran}
\bibliography{references_zotero}

\begin{IEEEbiography}[{\includegraphics[width=1in,height=1.25in,clip,keepaspectratio]{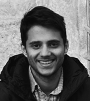}}]{A. Medina-Rull} received his M.S. degree in telecommunications engineering from University of Granada, Spain, in 2020. He is currently pursuing the Ph.D degree, focusing his research on 2D-materials based-device applications.
\end{IEEEbiography}

\begin{IEEEbiography}[{\includegraphics[width=1in,height=1.25in,clip,keepaspectratio]{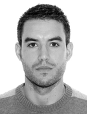}}]{F. Pasadas} received the Ph.D. degree in electronic and telecommunication engineering from the Universitat Aut\'onoma de Barcelona, Barcelona, Spain, in 2017. His current research interests include the modeling of 2D material-based devices and the design of novel radio-frequency applications based on such devices.
\end{IEEEbiography}

\begin{IEEEbiography}[{\includegraphics[width=1in,height=1.25in,clip,keepaspectratio]{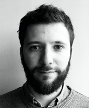}}]{E. G. Marin} received his Ph.D. degree in electronics from the University of Granada, Granada, Spain, in 2014. 

He has carried out his research career at IBM Research Zurich, Switzerland, at Cornell University, NY, USA and at University of Pisa, Italy. He is currently at Departament of Electronics at University of Granada. 
\end{IEEEbiography}

\begin{IEEEbiography}[{\includegraphics[width=1in,height=1.25in,clip,keepaspectratio]{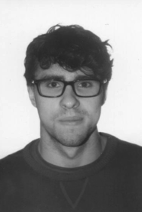}}]{A. Toral-Lopez} received his M.S. degree in telecommunications engineering from the University of Granada, Granada, Spain. He is currently pursuing the Ph.D degree in information and communication technologies at University of Granada. His research activity is focused in the simulation of 2D-materials based FET devices, targeting applications that range from radio-frequency to chemical and biochemical sensing.
\end{IEEEbiography}

\begin{IEEEbiography}[{\includegraphics[width=1in,height=1.25in,clip,keepaspectratio]{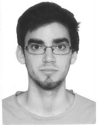}}]{J. Cuesta} received the B.S. degree in physics from the University of C\'ordoba, C\'ordoba, Spain, and the M.S. degree in physics from the University of Granada, Granada, Spain, where he is currently pursuing the Ph.D degree with a focus on simulation and modelling of 2D-materials based electronics and optoelectronics.
\end{IEEEbiography}

\begin{IEEEbiography}[{\includegraphics[width=1in,height=1.25in,clip,keepaspectratio]{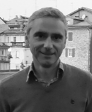}}]{A. Godoy} received the M.S. and Ph.D. degrees in physics from the University of Granada, Granada, Spain. He is currently Professor with the Department of Electronics, University of Granada, since 2011.
\end{IEEEbiography}

\begin{IEEEbiography}[{\includegraphics[width=1in,height=1.25in,clip,keepaspectratio]{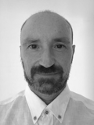}}]{D. Jim\'enez} received the Ph.D. degree in electronics engineering from the Universitat Aut\`onoma de Barcelona in 2000. 

He is currently Full Professor with the Departament d'Enginyeria Electr\`onica, Universitat  Aut\`onoma de Barcelona. He was a Visiting Researcher with the Universidad Aut\'onoma de Madrid in 2002; Universitat Rovira i Virgili in 2003; Tokyo Institute of Technology in 2009; Universidad de Granada in 2010; \'Ecole Polytechnique F\'ed\'erale de Lausanne in 2010; and the National Institute of Advance Science and Technology in 2013. His research activity has been focused on compact modeling of nanoscale transistors, including multiple-gate MOSFETs, silicon nanowire transistors, carbon nanotube transistors, and more recently graphene transistors. Other activities have been related with the research of new transistor architectures for low-power switching applications based on ferroelectric materials that exhibit negative capacitance. He has also been involved in the exploration of the ferroelectric resistive switching phenomenon targeting non-volatile memories. At the time being his main research is focused on 2D material based devices for both RF and digital applications.
\end{IEEEbiography}

\begin{IEEEbiography}[{\includegraphics[width=1in,height=1.25in,clip,keepaspectratio]{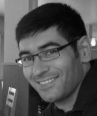}}]{F. G. Ruiz} received his telecommunication engineering degree from the University of M\'alaga, M\'alaga, Spain, in 2002, and his Ph.D. degree in physics from the University of Granada, Granada, Spain, in 2005. 

Since 2006, he is with the Department of Electronics of the University of Granada, where he co-founded the Pervasive Electronics Advanced Research Laboratory in 2018. He was a visiting researcher at TUDelft, UCL, IMEC, and Universitaet Siegen. His research focuses on 2D materials based electron devices, including RF devices and circuits, sensors, optoelectronic devices and neuromorphic devices and circuits.
\end{IEEEbiography}

\EOD

\end{document}